\documentclass[%
 reprint,
superscriptaddress,
nofootinbib,
 amsmath,amssymb,
 aps,
 pra, longbibliography,
]{revtex4-2}
\usepackage{layouts}
\usepackage{graphicx, subfigure}% Include figure files
\usepackage{dcolumn}% Align table columns on decimal point
\usepackage{bm}% bold math
\usepackage{mathtools}
\usepackage{amsmath}
\usepackage{pgfplots}
\usepackage{float}
\usepgfplotslibrary{fillbetween}
\pgfplotsset{compat=1.16}
\usetikzlibrary{fit,backgrounds,positioning,shapes,shapes.callouts}
    \usepackage[scaled=0.83]{helvet}
    \usepackage{marvosym,pifont}
    \usepackage[T1]{fontenc}
    \usepackage[utf8]{inputenc}
    \usepackage{ulem}
\usepackage{hyperref}

\begin{document}

%\preprint{APS/123-QED}

\title{Coherent interaction-free detection of noise}

\author{John J. McCord}
%\email[]{}

\affiliation{QTF  Centre  of  Excellence, 
 Department of Applied Physics, Aalto University, FI-00076 Aalto, Finland\\}
\affiliation{InstituteQ at Aalto University, FI-00076 Aalto, Finland\\}

\author{Shruti Dogra}

\affiliation{QTF  Centre  of  Excellence, 
	Department of Applied Physics, Aalto University, FI-00076 Aalto, Finland\\}
\affiliation{InstituteQ at Aalto University, FI-00076 Aalto, Finland\\}

\author{Gheorghe Sorin Paraoanu}

\affiliation{QTF  Centre  of  Excellence, 
	Department of Applied Physics, Aalto University, FI-00076 Aalto, Finland\\}
\affiliation{InstituteQ at Aalto University, FI-00076 Aalto, Finland\\}

\date{\today}% It is always \today, today,

\begin{abstract}

The measurement and characterization of noise is a flourishing area of research in mesoscopic physics. 
In this work, we propose interaction-free measurements as a noise-detection technique, exploring two conceptually different schemes: the coherent and the projective realizations. These detectors consist of a qutrit whose second transition is resonantly coupled to an oscillatory field that may have noise in amplitude or phase. 
For comparison, we consider a more standard detector previously discussed in this context: a qubit coupled in a similar way to the noise source. We find that the qutrit scheme offers clear advantages, allowing precise detection and characterization of the noise, while the qubit does not. Finally, we study the signature of noise correlations in the detector's signal.

\end{abstract}

\pacs{Valid PACS appear here}% PACS, the Physics and Astronomy
\maketitle

%\printinunitsof{in}\prntlen{\textwidth}

\section{Introduction}

The main obstacle in realizing large-scale quantum computers is noise, which hinders the realization of high-fidelity gates and readout \cite{Nielsen2010,Silveri_2017,Krantz2019}. This problem is often addressed with quantum error correction, which requires precise knowledge of the type of noise acting on the system, but may also be approached by passive methods such as dynamical decoupling, decoherence-free subspaces, and minimal noise subsystems \cite{Viola1999,Lidar2003,Wang2016}. Phase noise is also an important factor for another quantum technology: it affects the quantum bit error rate in cryptography protocols based on weak coherent states, for example, twin-field quantum key distribution \cite{Pirandola2020} that can, in principle, also be implemented in the microwave range. Thus, diagnosing various sources of noise and the errors they produce is of utmost importance for the success of fault-tolerant quantum computing  \cite{Martinis2015}. Noise is also a significant source of information for the dynamics of electrons at the nanoscale, as summarized by the famous dictum of Landauer, ``noise is the signal'' \cite{Landauer}.

Since qubits are highly sensitive to perturbations, a natural idea would be to use them as detectors of noise.  Indeed, in first-order perturbation theory, the excitation and decay probabilities are proportional to the noise spectral density at the negative and positive qubit frequencies, respectively \cite{Schoelkopf2003}. Alternatively, one can exploit the sensitivity to dephasing for magnetometry, where Ramsey interferometry with superconducting qubits has been used as a sensitive tool for measuring magnetic fields
\cite{Danilin2018,Blatter2018,Gusarov2023}. Several techniques have been proposed, such as using dynamical decoupling and its filtering properties to reconstruct the power spectral density \cite{Szankowski2017, Bylander2011},  employing the qubit as a vector network analyzer for characterizing the control lines \cite{Fedorov_2019}, and
identifying long-range correlations to reconstruct experimentally observed error rates \cite{Harper2020}. Further proposals include methods for characterizing low-frequency noise, where correlations can be obtained through repeated Ramsey measurements \cite{Dykman2023}, and using spectator qubits and machine learning to monitor noise in quantum processors \cite{Ferrie2023}.  In exploring the dynamics of electronic transport, significant effort has been dedicated to developing detectors sensitive to full counting statistics. Qubit-based detectors can be used to measure the characteristic function by performing Ramsey measurements at different values of the coupling \cite{Lebedev2016} or to extract the third cumulant from changes in their effective temperature \cite{Ojanen2007}. 

Here, we focus on the detection of oscillator noise, a paradigmatic type of noise which becomes relevant especially in quantum control -- when attempting to resonantly drive quantum systems which in general may interfere with the intended operations and lead to errors. We exploit a recent~\cite{dogra-ncomm-2022, jake-prr-2023} coherent interaction-free measurement (cIFM)~protocol for the detection of resonant noise in microwave circuits and investigate its efficacy at detecting both amplitude and phase noise. This scheme is based on a three-level quantum system (qutrit) whose basis states are labeled as ${\vert 0 \rangle, \vert 1 \rangle, \vert 2 \rangle }$, where the allowed transition between levels $\vert 0 \rangle - \vert 1 \rangle$ and levels $\vert 1 \rangle - \vert 2 \rangle$ corresponds to transition frequencies $\nu_{01}$ and $\nu_{12}$, respectively.
As per the cIFM protocol, there is a train of identical beam-splitter unitaries targeting the $\vert 0 \rangle - \vert 1 \rangle$ transition, with its consecutive blocks being separated by a fixed duration. In between each pair of beam-splitter unitaries, $\vert 1 \rangle - \vert 2 \rangle$ microwave pulses called $B$ pulses may be sandwiched; whose presence is ascertained in an interaction-free manner~\cite{dogra-ncomm-2022, jake-prr-2023}. There are three possible outcomes of the protocol which leave the three-level system in one of the basis states ($\vert 0 \rangle, \vert 1 \rangle, \vert 2 \rangle$) with respective occupation probabilities: $p_0$, $p_1$, and $p_2$. For a qutrit initialized in its ground state $\vert 0 \rangle$, and undergoing the cIFM protocol, one can have a successful interaction-free detection of a $B$ pulse with probability $p_0$, a non-desirable non-interaction-free detection with a probability $p_2$, and inconclusive results with probability $p_1$. These probabilities have a direct correspondence with the populations of the respective energy levels of the qutrit. A different interaction-free concept, which we call projective interaction-free measurement (pIFM), interjects projective measurements on state $|2\rangle$ after each interaction with the microwave $B$ pulses \cite{jake-prr-2023}. 
Projective interaction-free measurements have been performed in various quantum optics experiments that followed the original theoretical proposal \cite{Kwiat_1995, Kwiat_1999,Ma_2014,Peise_2015}. The projective measurement needed can also be implemented in circuit quantum electrodynamics, for example by employing the switching of a Josephson junction when one of the excited states in the washboard potential is close to being delocalized \cite{Paraoanu_2006,Paraoanu_2011,Paraoanu_2011_2}. 
It has also been proposed to use the Zeeman states of a trapped ion in conjunction with polarized photon states as a means of realizing projective interaction-free measurements \cite{Pavicic2007}.
For non-random pulses, the coherent protocol turns out to be more efficient. In fact, it has been shown that the coherent protocol reaches the Heisenberg limit when the Fisher information is evaluated at small strengths of the $B$ pulses, whereas the projective protocol only reaches the standard quantum limit \cite{jake-prr-2023}.

We study noise detection using the cIFM and pIFM protocols in a systematic manner, by considering a drive acting resonantly on the  $\vert 1 \rangle - \vert 2 \rangle$ transition. Noise can be present either in the amplitude or in the phase of the drive. If the correlation time of the noise is much larger than the total duration $T$ of the sequence plus the measurement time, the problem of characterizing the noise is trivial, since each nearly-constant value of the drive can be detected with high efficiency. The interesting situation that we consider in this work, is when the correlation time is much larger than $\tau_{\rm B}$ and of the same order or smaller than $T$. This allows us to sample the noise in small $\tau_{\rm B}$ intervals where it is nearly constant. This arrangement requires, ideally, that $N$ is very large, while in real experiments $N$ is limited by decoherence.

To understand the advantage of interaction-free measurements, we consider for comparison a paradigmatic detector based on absorption, consisting of a single qubit with transition $\vert \rm{g} \rangle - \vert \rm{e} \rangle$  at the frequency $\omega_{\rm ge}$, which interacts resonantly with the noise. The simplest detection scheme is to allow a qubit  to evolve under this noise and read the qubit's state after some time. If the noise has reasonably strong coupling with the detector qubit, then the state of the qubit will be influenced in the presence of noise. Therefore, a qubit initialized in its ground state $\vert {\rm g} \rangle$ exhibits non-zero probability to be found in the excited state $\vert \rm{e} \rangle$. Consequently, one can use the excited state population $p_{\rm e}$ as a marker to ascertain the presence of noise. This mechanism might seem simple and useful at first, but this is not so reliable in practice. The detector qubit evolves randomly under the influence of this noise leading to arbitrarily varying outcomes that average to zero. Moreover, if the noise sums arbitrarily close to zero in a given time, the qubit detector will not be able to detect the noise.

The paper is organized as follows: In Sec. \ref{models}, we introduce the three detector models: the qubit and the two interaction-free protocols utilizing the qutrit. Our main results are presented in Sec. \ref{results}, where we consider (white) noise with a small correlation time relative to the total duration $T$, yielding results consistent with the standard decoherence approach. Sec. \ref{binary_noise} discusses the case of binary noise described by a Poisson probability distribution with a correlation time comparable to $T$. In Sec. \ref{noise_correlations}, we examine the signatures of autocorrelations in the detector output, again with a correlation time comparable to $T$. Then, in Sec. \ref{appl}, we present two experimental platforms -- the flux qutrit and Rydberg atoms -- where our protocols can be readily implemented. We conclude in Sec. \ref{conclusions}.

\section{Detector models \label{models}}
In the following subsections, we describe systematic and efficient techniques to detect resonant noise, exploiting qutrit-based protocols. Further, we compare the efficacies of these qubit-based and qutrit-based models to detect noise, highlighting the difference between absorptive and interaction-free measurements.
In both cases we start with a generic oscillatory noisy source %$\Omega (t) = \Omega_{0}(t) \cos [\omega_{0}t + \chi (t)]$
at a frequency $\omega_{0}$, which is resonantly coupled into the corresponding transition with a generic Rabi coupling. %$\Omega_{0}$.
The phase $\chi (t)$ is in general noisy, and we can also separate a noisy amplitude component  $\zeta (t)$ in the Rabi coupling. % $\Omega_{0} (t) = \Omega_{0} + \zeta (t)$.
An overview of standard notations and results related to amplitude and phase noise is presented in Appendix A.
As we shall see, successful detection is established when the population $p_{\rm e}$ on the excited state $|{\rm e}\rangle$ for the qubit or the population $p_{0}$ of the ground state $|0\rangle$ for the qutrit is nearly 1. Finding the detector in these respective states is therefore highly indicative of the presence of noise. We will refer to these probabilities generically as {\it marker populations}. The occupation probabilities on either of these states can be obtained by partial tomography, depending on the specific experimental platform (see, {\it e.g.}, Sec. \ref{appl} for some specific examples).

\begin{figure}[h]
	\centering
	\includegraphics[scale = 0.373,keepaspectratio=true]{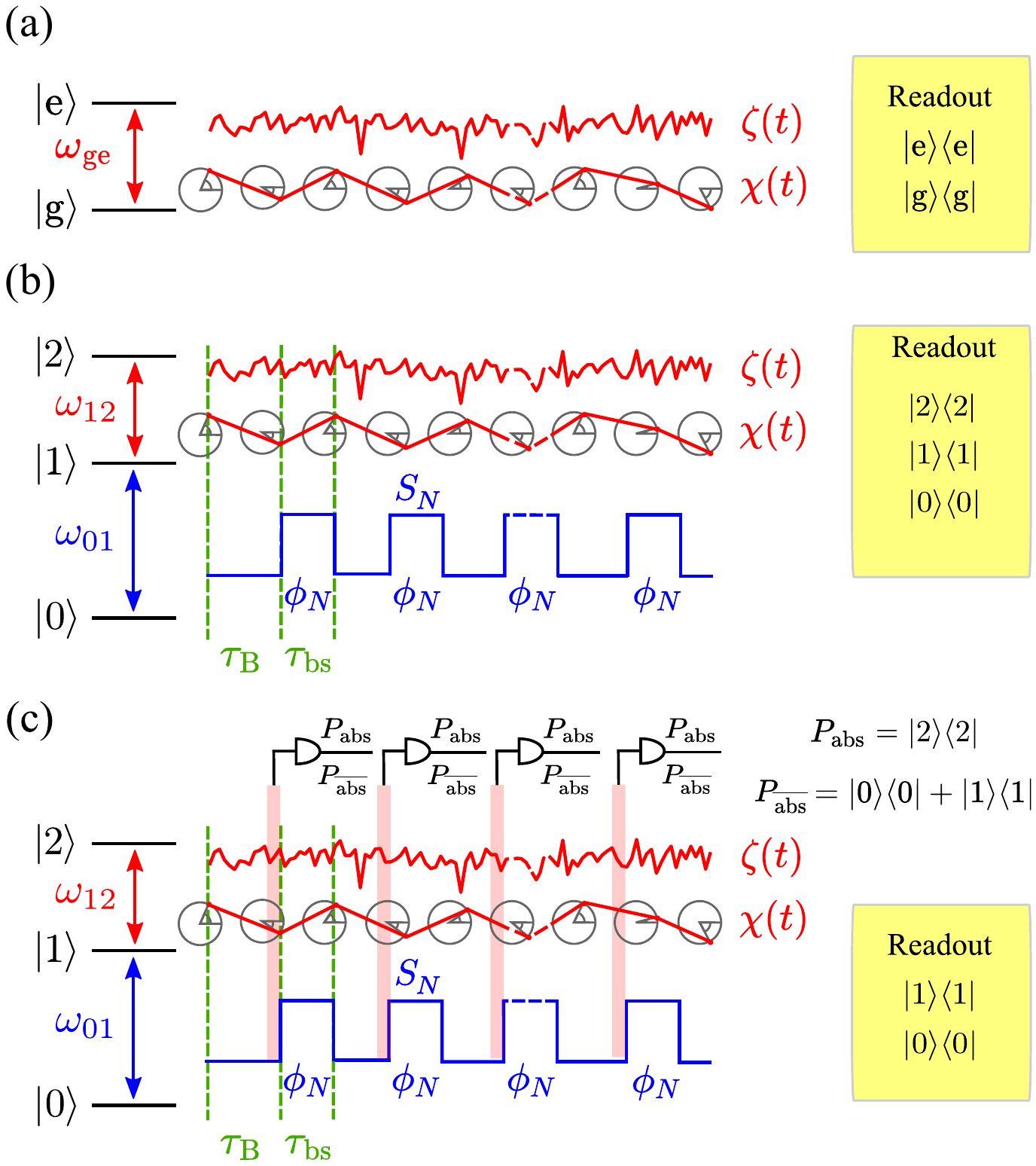}
	\caption{The three noise detection schemes studied in this work: (a) qubit, (b) cIFM, and (c) pIFM. The qubit detector is an aborptive detector, whereas cIFM and pIFM detectors utilize interaction-free measurements on a qutrit by employing a sequence of Ramsey pulses on the $|0\rangle - |1\rangle$ transition. Noise is coupled into the $|\mathrm{g}\rangle - |\mathrm{e}\rangle$ transition in the case of the qubit detector and into the $|1\rangle -|2\rangle$ transition for the qutrit. In the case of the pIFM, the unitary evolution is interrupted by a detector that is triggered if the state of the qutrit is $|2\rangle$ and does not produce a detection event otherwise. Finally, at time $T$, a partial tomography (population detection) is performed at the end of the sequence.} 
	\label{detection_schemes}
\end{figure}

\subsection{Qubit-based detector}

Consider a qubit with the computational basis denoted by ground and excited states $|{\rm g}\rangle $, $|{\rm e} \rangle$, see Fig. \ref{detection_schemes}(a). The Hamiltonian under the drive provided by the noisy oscillator is 
	\begin{equation}
		H_{\rm ge} = \hbar \omega_{\rm ge} |{\rm e}\rangle \langle {\rm e}| + \hbar \Omega_{\rm ge} (t)\cos (\omega_{0} t + \chi(t))
		[|{\rm e}\rangle \langle {\rm g}| + |{\rm g}\rangle \langle {\rm e}|],
	\end{equation}
	where $\Omega_{\rm ge}(t) = \Omega_{\rm ge} + \zeta (t)$, and $\zeta (t)$ is the amplitude noise.

By introducing a unitary $U_{\rm ge} = |{\rm g}\rangle \langle {\rm g}| + e^{i \omega_{\rm ge}t} |{\rm e}\rangle \langle {\rm e}|$ we can transform this Hamiltonian into a frame rotating at the qubit frequency,   $H_{\rm ge} \rightarrow U_{\rm ge} H_{\rm ge}
U_{\rm ge}^{\dag} + i \hbar (dU_{\rm ge}/dt)U_{\rm ge}^{\dag}$,  obtaining in the rotating wave approximation and at resonance ($\omega_{\rm ge} = \omega_{0}$),

	\begin{eqnarray}
		H_{\rm ge}(t) &=& \frac{\hbar \Omega_{\rm ge}(t)}{2}	\left[ e^{- i \chi(t)} |{\rm e}\rangle \langle {\rm g}| +  e^{i \chi (t)} |{\rm g}\rangle \langle {\rm e}| \right] \\ 
		&=& \frac{\hbar \Omega_{\rm ge}(t)\cos \chi (t)}{2} \sigma_{\rm ge}^{x} - \frac{\hbar \Omega_{\rm ge}(t)\sin \chi (t)}{2} \sigma_{\rm ge}^{y} \\
		&=& \frac{\hbar \Omega_{\rm ge}(t)}{2}\hat{\boldsymbol{n}}_{\chi}(t) \cdot \boldsymbol{\sigma}_{\rm ge},
	\end{eqnarray}
 where $\sigma_{\rm ge}^{x} = |{\rm g}\rangle \langle {\rm e}| + |{\rm e}\rangle \langle {\rm g}|$ and $\sigma_{\rm ge}^{y} = -i |{\rm g}\rangle \langle {\rm e}| + i|{\rm e}\rangle \langle {\rm g}|$, $\boldsymbol{\sigma}_{\rm ge} = (\sigma_{\rm ge}^{x}, \sigma_{\rm ge}^{y})$, and $\hat{\boldsymbol{n}}_{\chi}(t) = (\cos \chi (t), - \sin \chi (t))$
is a rotation axis in the $xOy$ plane. In general, the Hamiltonian above does not commute with itself at different times. To deal with this issue, we divide the time into $N$ intervals $j$ of duration $\tau_{\rm B}$, during which $\chi (t)$ is approximately constant. In this case, the phase $\varphi$ of the unitary transformation is the same as the noise phase $\chi(t)$. 
During these intervals, the unitary transformation produced by the pulses is 

\begin{equation}
	B_{\rm ge} (\theta_{j},\varphi_{j}) = e^{-i \theta_{j}\hat{\boldsymbol{n}}_{j}\cdot\boldsymbol{\sigma}_{\rm ge}/2} = \mathbb{I}_{\rm ge} \cos \frac{\theta_j}{2} - i (\hat{\boldsymbol{n}}_{j}\cdot\boldsymbol{\sigma}_{\rm ge})\sin \frac{\theta_j}{2}, \label{eq:bpls}
\end{equation}
where 
	$\theta_{j} = \int_{t_j}^{t_j + \tau_{\rm B}} \Omega_{\rm ge}(t) dt = \Omega_{\rm ge}\tau_{\rm B} + \int_{t_j}^{t_j + \tau_{\rm B}} \zeta (t) dt$ is the arbitrary angle corresponding to the noisy drive~\cite{bell-1974}, $\hat{\boldsymbol{n}}_j = (\cos \varphi_j , - \sin \varphi_j)$ is the axis of rotation, and $\mathbb{I}_{\rm ge}$ is the unit $2\times 2$ matrix. Here, $t_j$ and $t_j + \tau_{\rm B}$ are the initial and final times of the intervals. 

In a more general situation the noise phase $\chi(t)$ varies significantly; in this case the unitary transformation of duration $\tau_{\rm B}$, effective angle $\theta_j$, and an overall axis of rotation $\varphi_j$ can be written as 
\begin{equation}
	B_{\rm ge}(\theta_j,\varphi_j) = e^{-i \theta_{j}\hat{\boldsymbol{n}}_{j} \cdot \boldsymbol{\sigma}_{\rm ge}/2}  = \prod_{p=1}^{\mathcal{P}} e^{-i \delta \theta_{p} \hat{\boldsymbol{n}}_{\chi_{p}} \cdot \boldsymbol{\sigma}_{\rm ge}/2},  \label{eq:bj}
\end{equation}
where $\delta \theta_p = \Omega_{\rm ge}(t) \delta t$ is the effective angle of rotation along the axis $\hat{\boldsymbol{n}}_{\chi_{p}}(t) = (\cos \chi_{p} (t), - \sin \chi_{p} (t))$ during the ${p}$th transient of duration $\delta t$. Here, $\delta t$ is the infinitesimal time interval during which the noise amplitude $\zeta(t)$ and the noise phase $\chi(t)$ are approximately constant, which in the worst case is the inverse of the noise sampling rate. The number of noise samples in duration $\tau_{\rm B}$ is denoted by $\mathcal{P}$, which is approximately equal to the ratio $\tau_{\rm B}/(\delta t)$.

\subsection{Qutrit-based detectors} 

Our models to detect noise using a qutrit with computational basis states  $(|0\rangle, |1\rangle, |2\rangle)$ are based on the cIFM and pIFM  protocols, which aim to efficiently detect noise resonant with the  $\vert 1 \rangle - \vert 2 \rangle$ transition.
A crucial component of these protocols is the implementation of additional beam-splitter pulses of duration $\tau_{\rm bs}$, which are realized by resonantly coupling a control field into the $\vert 0 \rangle - \vert 1 \rangle$ transition, as shown in Fig. \ref{detection_schemes} (b)(c).
The Hamiltonian under these drives is 

\begin{eqnarray}
		H &=& \hbar \omega_{01} |1\rangle \langle 1| + \hbar (\omega_{01} +\omega_{12}) |2\rangle \langle 2|  \nonumber \\  
		&+& \hbar \Omega_{01} (t)\cos (\omega_{01} t)
		[|1\rangle \langle 0| + |0\rangle \langle 1|] \nonumber \\
		&+&  \hbar \Omega_{12} (t)\cos (\omega_{0} t + \chi(t))
		[|2\rangle \langle 1| + |1\rangle \langle 2|],
	\end{eqnarray}
where $\Omega_{12}(t) = \Omega_{12}+ \zeta(t)$ consists of $\zeta(t)$, the noisy part in the amplitude of the field coupled to the $|1\rangle - |2\rangle$ transition. This amplitude noise is shown as the red-colored arbitrarily varying signal in each of the protocols illustrated in Fig.\ref{detection_schemes}. The phase noise $\chi(t)$ is also depicted as a red signal in the schematic of each protocol.

With the unitary $U = |0\rangle \langle 0| + e^{i \omega_{01}t} |1\rangle \langle 1| +
e^{i (\omega_{01}+ \omega_{12})t} |2\rangle \langle 2|$ we can transform this Hamiltonian as $H \rightarrow U H
U^{\dag} + i \hbar \frac{dU}{dt}U^{\dag}$ and apply the rotating wave approximation under the resonance condition $\omega_{0}=\omega_{12}$ to obtain:
	\begin{eqnarray}
		H(t) &=& \frac{i\hbar \Omega_{01}(t)}{2} [|1\rangle \langle 0| - |0\rangle \langle 1|] \nonumber \\
		&+&  \frac{\hbar \Omega_{12}(t)}{2} [e^{-i\chi (t)} |2\rangle \langle 1| + e^{i\chi (t)}|1\rangle \langle 2|]. 
	\end{eqnarray}

The cIFM and pIFM protocols employ a series of beam-splitter pulses of duration $\tau_{\rm bs}$ on the  $\vert 0 \rangle - \vert 1 \rangle$ transition, intercalated with detection times $\tau_{\rm B}$ onto which the noise is sensed.
We denote $\mathbb{I}_{kl} = |k\rangle \langle k| + |l\rangle \langle l|$, $\sigma^{y}_{kl} = -i|k\rangle \langle l| + i|l\rangle \langle k|$, $\sigma^{x}_{kl}= |k \rangle \langle l| + |l\rangle \langle k|$, with $k,l \in \{0,1,2\}$ and $k < l$ 
that are described by the unitary 
\begin{eqnarray}
	S ( \phi_{N} ) &=& e^{-i\phi_{N} \sigma^{y}_{01}/2} \\
	& & = \mathbb{I}_{01} \cos\frac{\phi_{N}}{2} - i \sigma_{01}^{y} \sin\frac{\phi_{N}}{2} + |2\rangle\langle 2|.
\end{eqnarray}
Here, the beam-splitter strengths $\phi_{N}$ are chosen such that 
$\phi_{N} = \pi/(N+1)$ by appropriately choosing the Rabi strengths 
$\phi_{N} = \int \Omega_{01}(t) dt$ corresponding to each pulse.
We use similar notations as for the qubit detector,
$\hat{\boldsymbol{n}}_{j} = (\cos\varphi_j, -\sin\varphi_j )$, when $\chi(t)$ is approximately constant for the duration $\tau_{\rm B}$, i.e., $\chi(t) = \varphi(t)$, and $\boldsymbol{\sigma}_{12}=
(\sigma_{12}^{x}, \sigma_{12}^{y})$. Explicitly, the unitary operation $B (\theta_{j},\varphi_j )$ is given by
\begin{eqnarray}
	B (\theta_{j},\varphi_j ) &=& e^{-i \theta_{j}\hat{\boldsymbol{n}}_{j}\cdot\boldsymbol{\sigma}_{12}/2} \\ 
	&=& |0\rangle \langle 0| + \mathbb{I}_{12} \cos \frac{\theta_j}{2} - i (\hat{\boldsymbol{n}}_{j}\cdot\boldsymbol{\sigma}_{12})\sin \frac{\theta_j}{2}, 
\end{eqnarray}
where the definition of $\theta_j$ used in the qubit case applies. If $\chi(t)$ is not constant, the effective unitary transformation over the duration $\tau_{\rm B}$ takes a form similar to Eq.~\ref{eq:bj}, 
\begin{equation}
	B(\theta_j, \varphi_j) = \prod_{p=1}^{\mathcal{P}} e^{-i \delta \theta_{p} \hat{\boldsymbol{n}}_{\chi_{p}}\cdot\boldsymbol{\sigma}_{12}/2}.  \label{Eq:16}
\end{equation}

\paragraph{Coherent IFM (cIFM) - based protocol}

The cIFM protocol involves using a train of beam-splitter unitaries $S (\phi_N )$ with a duration $\tau_{\rm bs}$, separated from each other by an interval $\tau_{\rm B}$, as shown in  Fig. \ref{detection_schemes}(b). These are applied resonantly to the first transition, while the noise couples into the second transition.
We consider a sample of this noise for a duration $T = (N+1)(\tau_{\rm bs} + \tau_{\rm B})$, initialize our detector (qutrit) in state $\vert 0 \rangle$, and allow it to evolve with the series of beam-splitter unitaries. Results from this protocol are read in a counter-intuitive manner; i.e., if no noise is present, the qutrit is found in state $\vert 1 \rangle$, while in the presence of noise, the state of the qutrit remains the same (ground state $\vert 0 \rangle$) with high probability. We use the
ground state probability $p_0$ of the qutrit as a marker for the detection of noise. We obtain $p_0$ values at time $T$ from several implementations with $N \in \{1, ..., 100\}$. The whole process is then repeated several times, and the average value of $p_0$, i.e., E[$p_0$] is observed.

\paragraph{Projective IFM (pIFM) - based model}

This is also a qutrit-based model to detect resonant noise, which we present schematically in Fig. \ref{detection_schemes}(c). As described earlier, in the pIFM-based model, there are also $(N+1)$ beam-splitter unitaries of duration $\tau_{\rm bs}$, each implementing a rotation of angle $\phi_{N}=\pi/(N+1)$ around the $y$ axis. Similar to the cIFM protocol, the noise acts at the frequency $\omega_{12}$. Unlike the cIFM protocol, where coherences are preserved as an asset to be used later, in pIFM, coherences between levels $\vert 1 \rangle - \vert 2 \rangle$ are erased via projective measurements at the end of each noise pulse interaction with the detector, i.e., at
times $j(\tau_{\rm bs} + \tau_{\rm B})$, where $j\in \{1, ..., N\}$.
These projectors, which are applied immediately after each noise pulse, are defined as $P_{\rm{abs}} = \vert 2 \rangle \langle 2 \vert$ (detection of excitation on $|2\rangle$) and $P_{\overline{\rm{abs}}} = \vert 0\rangle \langle 0 \vert + \vert 1 \rangle \langle 1 \vert$ (absence of a detection event on $|2\rangle$). Here, we also use the ground state population as a marker, a non-zero value which is the signature of noise.

\section{Detection of white noise \label{results}}

In this section, we consider noise with correlation times much smaller than $T$, such that different noise events are almost independent of each other and are hence uncorrelated. This noise can be assumed to be effectively white without loss of generality. Further,
	we allow the qubit/qutrit detectors to interact with noise for a fixed amount of time and measure their respective final states. This process is repeated several times and the final qubit/qutrit states obtained are averaged out. This leads to the same results as expected from the standard master equation approach, where correlations are neglected with respect to the time $T$ \cite{ Schoelkopf2003, Preskill2018, Krantz2019}.

We simulate qubit-based and qutrit-based detectors to ascertain the presence of resonant white Gaussian noise with a maximum amplitude $|\zeta(t)|_{\rm max} = \max (\theta_j)/\tau_{\rm B}$, 
%\sout{We consider different scenarios and compare the corresponding results. Moreover, we} 
and we analyze the detection in three possible situations: (i) variation of $\zeta(t)$ at a constant phase, i.e., amplitude noise, (ii) variation of $\chi(t)$ with $\zeta(t)$ constant in time, i.e., phase noise, and (iii) a general case of both $\zeta(t)$ and $\chi(t)$ varying with time, i.e., amplitude and phase noise. For concreteness, in superconducting-circuit-based realizations, we could have a sampling rate of $10^7$ samples/s, as well as beam-splitter and sensing times of $\tau_{\rm bs} = 20$ ns and $\tau_{\rm B} = 200$ ns, respectively. In all these cases, %the noise is targeting only one specific frequency $\omega_{\rm ge}$ (for the qubit case) or $\omega_{12}$ (for the qutrit case) and 
we divide this noise into several consecutive intervals of length $\tau_{\rm B}$ and $\tau_{\rm bs}$.

The evolution in the  $j$th interval can be described by a unitary pulse $B(\theta_j)$ of duration $\tau_{\rm B}$, with an effective angle $\theta_j$ and an overall axis of rotation $\varphi_j$.
We assume that in the cIFM and pIFM protocols, the three-level quantum system undergoes nearly instantaneous beam-splitter operations, as ensured by the condition  $\tau_{\rm bs} << \tau_{\rm B}$. This produces a negligible error in the case of continuous noise, where $\tau_{\rm bs}$ is the time for which there exists simultaneous driving of $\vert 0 \rangle - \vert 1 \rangle$ and $\vert 1 \rangle - \vert 2 \rangle$. Thus, the sequence can be simplified to a series of beam-splitter unitaries and unitary pulses of arbitrary angles $\theta_j$.

\subsubsection{Amplitude noise}

\begin{figure*}[ht]
%\begin{figure}
	\centering
	\includegraphics[width=0.9\linewidth, angle=0]{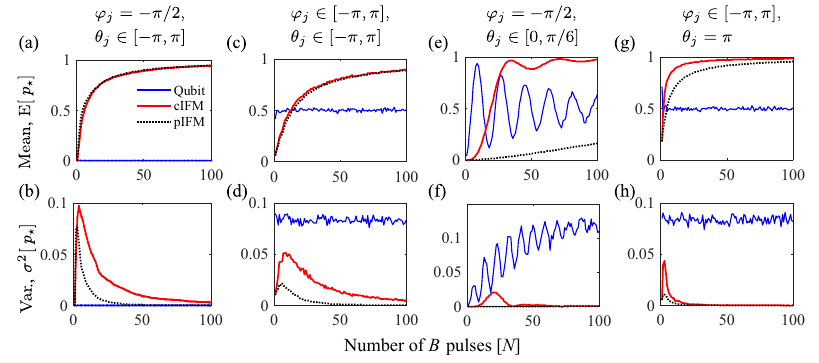}
	\caption{ Mean (top row) and variance (bottom row) of the marker populations ($p_{\star}$), i.e., $p_{\rm e}$ for the qubit-based detector and $p_0$ for the qutrit-based detectors -- extracted from 500 realizations of the protocol, each with $\Omega_{\rm{ge}(12)} = 0$, and $N \in \{1,...,100 \}$. Panels (a) and (b) correspond to the case of amplitude noise with a net sum of zero, i.e., $\sum_{j=1}^N \theta_j = 0$. Results from the general case are shown in panels (c) and (d), which include both amplitude and phase noise. Panels (e) and (f) correspond to the case of amplitude noise with small values of arbitrarily chosen $\theta$. Panels (g) and (h) show the results for phase noise at a constant noise amplitude. The ranges of amplitude ($\theta$) and phase ($\varphi$) are given at the top of each column. All noises considered are white Gaussian.
	}
	\label{figure2}
\end{figure*}

We first consider amplitude noise, which in each interval $j$ produces a unitary pulse $B$ of duration $\tau_{\rm B}$ and effective angle 
$\theta_j = \Omega_{\textrm{ge} (12)} \tau_{\rm B} + \int_{t_j}^{t_j + \tau_{\rm B}}\zeta(t) dt$ with a fixed axis of rotation ($\chi = -\pi/2$ and hence, $\varphi_j = -\pi/2$).  Here, $t_j = j\tau_{\rm bs} + (j-1)\tau_{\rm B}$ and $t_j + \tau_{\rm B} = j (\tau_{\rm B} +\tau_{\rm bs})$ are the initial and final times of each pulse, with $j\in \{1, ..., N\}$.
To clearly demonstrate the difference between qubit and qutrit detectors, we engineer the noise at a sampling rate of $5\times 10^6$ samples/s, ensuring its net sum over a long period is arbitrarily close to zero with a signal-to-noise ratio (SNR) of 1. Specifically, in this case, we have $\sum_{j=1}^N \theta_j = 0$, with a constant noise amplitude during a given $B$ pulse duration, such that $\theta_j = \zeta(t_j) \tau_{\rm B}$. The results from this simple model are shown in Fig.~\ref{figure2}(a,b). In  Fig.~\ref{figure2}(a), we present the mean value of the marker populations (E[$p_{\rm e}$] for the qubit and E[$p_0$] for the qutrit), averaged over $500$ realizations of the same experiment for various values of $N \in \{1, ..., 100\}$.
 Fig.~\ref{figure2}(b) presents the corresponding variance values for this state. Here, the continuous blue curve represents the excited state population of the qubit-based detector, which is nearly zero; therefore, the qubit detector completely misses the presence of noise. Further, the continuous red curve and the dashed black curve correspond to the average value of $p_0$ resulting from cIFM and pIFM protocols, respectively. In both cases, E[$p_0$] approaches 1 for large $N$, signifying that both cIFM and pIFM-based detectors are almost equally efficient at detecting noise in such scenarios.

In general, the net sum of the noise may not approach zero over a long time range ($\approx T$). In that case, the qubit detector will evolve with the net sum of the noise, such that $p_{\rm e} = \sin^2 (\theta_{T})$, where $\theta_{T} = \sum_{j=1}^{N} \theta_j $. 
Thus, the mean value E$[p_{\rm e}]$ approaches $0.5$ after several repetitions, which is also consistent with the average value of $\sin^2 (\theta_{T})$ for $\theta_{T} \in [0, \pi]$. 
Such situations are shown in  Fig.~\ref{figure2}(c,e,g), and are discussed in the following subsections.

We also consider a situation with only positive values of noise, i.e., $\zeta(t) > 0$, and observe that the qubit detector leads to the same outcome, as expected. Interestingly, the pIFM-based qutrit detector also yields the same outcomes, while the cIFM protocol leads to improvement in the average values. 
In this case, the cIFM protocol outperforms the pIFM protocol for the detection of positive amplitude noise. In special circumstances, where the noise sampling rate $=\tau_{\rm B}^{-1}$, such that there is only one noise sample ($\mathcal{P}=1$) in the entire $\tau_{\rm B}$ duration, the pIFM protocol is independent of the axis of rotation $\varphi_j$, while the cIFM protocol is very sensitive to it. Thus, we can acquire information about the phase $\varphi_j$ from the cIFM protocol but not from the pIFM protocol.
Next, we consider a situation with small values of $\theta_j \in [0, \pi/6]$ with $\varphi_j=-\pi/2$, as shown in Fig.~\ref{figure2}(e,f). In this case, E[$p_0$] from cIFM approaches $1$ for $N>20$, which is much better than pIFM, where E[$p_0] \approx 0.25$ for $N=100$. The qubit initially oscillates at $\sin^2(\sum \theta_j)$ and finally attains the value $0.5$.

\subsubsection{Amplitude and phase noise}

A more general noise may have time dependence for both its amplitude and phase. The results from this general scenario are shown in Fig.~\ref{figure2}(c,d). As expected, the mean values for the qubit detector tend to stay close to $0.5$. For strong enough noise, such that $\theta_{j} \in [0,\pi]$, both mean and variance values are independent of the value of $N$. Thus, by increasing the value of $N$, i.e., for a larger $T$, we do not see any enhancement in the detection of noise with this absorption-based qubit detector. The best result that this detection protocol can yield in this case is the maximally mixed state of the qubit, leading to an equally populated ground state and excited state. This is equivalent to obtaining the mean values of the populations, ${\rm E}[p_{\rm g}] = {\rm E}[p_{\rm e}] = 0.5$ with significantly large values of variance. Thus, we can conclude that due to quite large variance values, a widely varying output, and having less sensitivity, the qubit detector is less efficient.

We then allow the same noise to be accessed by the qutrit detectors, and the corresponding mean and variance values are shown as the continuous red curve for the case of cIFM and as the dashed black curve for the case of pIFM in Fig.~\ref{figure2}. For large values of $N$, the variance is quite close to zero and E[$p_0$] is close to 1, signifying a very efficient detection of noise. Interestingly, the continuous red and dashed black curves in Fig.~\ref{figure2}(c,d) follow a similar trend as those in Fig.~\ref{figure2}(a,b). This demonstrates the efficiency of qutrit-based protocols irrespective of whether noise sums to zero or not.

 \subsubsection{Phase noise}
 
 Here, we consider a constant amplitude of noise such that $\zeta(t) \propto \pi/\tau_{\rm B}$ and an arbitrarily chosen phase, $\varphi_{j} \in [-\pi, \pi]$, with a noise sampling rate of $10^7$ samples/s, resulting in two noise samples in the $j$th pulse, $\mathcal{P}=2$. In the case of phase noise, for situations with $\mathcal{P} > 1$, $\theta_j$ values may differ for different $j$ as per Eq.~\ref{Eq:16}, even if the noise amplitude $\zeta(t)$ is constant. The corresponding results are shown in Fig.~\ref{figure2}(g,h).
As expected, the pIFM-based protocol is less sensitive to changes in $\varphi_{j}$. However, the cIFM-based protocol is highly sensitive to variations in $\varphi_{j}$ and can thus be more effective at determining the nature of the noise. Moreover, for $\mathcal{P}=1$, the pIFM-based protocol is not sensitive to changes in $\varphi_{j}$ and hence cannot characterize phase noise. The qubit-based protocol is the least informative about noise, with its mean value staying close to $0.5$ with significantly high values of variance. Additionally, the qubit-based protocol does not detect the presence of phase noise when $\theta$ is an integral multiple of $\pi$.

\section{Detection of binary processes \label{binary_noise}}

In this and the following sections, we consider noise correlation times on the order of $T$. Specifically, we focus on binary noise, e.g., generation-recombination noise and random telegraph (burst) noise, which span the correlation times in a wide range from $T/100$ to $T$, and attempt to detect its presence via IFM-based protocols \cite{Yamamoto-2004, Balandin2013, Balandin2024, Kirton1989}.
We model the noise using a Poisson point process, which results in noise with steps of $\pm \pi$:

\begin{equation}
	P (m,T) = \frac{(\kappa T)^m}{m!} e^{-\kappa T}, \label{eq:Poisson}
\end{equation} where $\kappa$ is the switching frequency and $P(m,T)$ represents the probability of $m$ switching events during the time interval $T$. This process has an exponentially-decaying autocorrelation function and a Lorentzian power spectral density. 
 Poissonian processes are fundamentally important because they are simple and can be used as building blocks for generating processes with power-law spectral densities by considering that the time $\kappa^{-1}$ is probabilistically distributed (see Appendix A) \cite{Ziel1979,Dutta1981,Paladino2014}. 

\begin{figure}
	\centering
\includegraphics[width=0.9\linewidth]{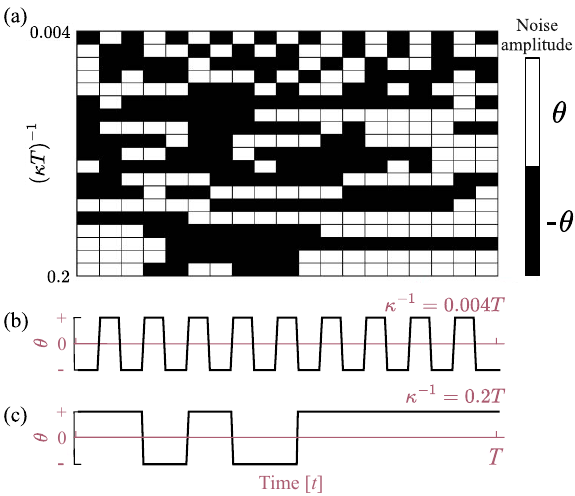}
  \caption{Binary noise over time $T$, generated by a Poisson point process with mean $\langle m\rangle = \kappa T$ and correlation time  $\kappa^{-1} \in [T/250, T/5]$. Panel (a) displays a complete matrix representation of one realization of this noise, while panels (b) and (c) show 1D traces corresponding to two extreme values of the correlation time.
	}
\label{fig_noise}
\end{figure}

The correlation time $\kappa^{-1}$ is considered such that $ T \geq \kappa^{-1} > 0$, where $T = (N+1)(\tau_{\rm B} + \tau_{\rm bs})$. The mean $\langle m\rangle$ of the distribution $P(m,T)$ is $\langle m\rangle = \kappa T$. Thus, as %$\tau$ increases
$\kappa$ decreases, the switching frequency decreases, leading to a decrease in the mean and variance of the distribution. Fig.~\ref{fig_noise} shows an example of noise with amplitudes $\pm \theta$ where the phase can be flipped at a rate of up to $20/T$ for the duration $T$. 
This allows for a maximum of $20$ switching events or noise samples within $T$,
with $\kappa^{-1}$ varying linearly from $\tau_{\rm bs}$ to $T$. To enhance clarity, Fig.~\ref{fig_noise} shows only a part of this noise, with values of $\kappa^{-1}$ being limited to the range $\kappa^{-1} \in [\tau_{\rm bs}, 0.2T]$.
A qubit detector would be very inefficient at detecting this type of noise. An intuitive explanation for this is given in Appendix B.

Here, we first analyze the case of binary noise, switching between $\pm \theta$ at a rate of up to $10^9$ times in one second. We consider an intercept of such a noise for a fixed duration of time $T$ and try to detect it using cIFM and pIFM protocols. Fixing $T$ and taking $\tau_{\rm bs} = T/400$ or as small as possible, we arbitrarily choose the value of $N \in \{1,\dots, 40\}$. For instance, $N=1$ requires two beam-splitter unitaries $S_{1}(\pi/2)$: one at the start and one at the end of the noise, with $B$ pulse duration $\tau_{\rm B} = T$. For any $N$, $N+1$ beam-splitter unitaries $S(\phi_{N})$ are placed at intervals of $\tau_{\rm B} = T/N$ on resonance with the $|0\rangle-|1\rangle$ transition frequency, with the noise coupled as before into the $|1\rangle - |2\rangle$ transition. Ideally, the protocol is designed in such a way that the beam-splitter pulses act instantaneously with $\tau_{\rm bs} \rightarrow 0$. However, due to the constraints set by the quantum speed limit and experimental feasibility, $\tau_{\rm bs}$ is finite. The values of $T$, $\tau_{\rm bs}$, and $N$ are chosen such that even for the largest $N$, $\tau_{\rm bs} << \tau_{\rm B}$, and the qutrit's evolution under the $|1\rangle - |2\rangle$ drive can be ignored during the short intervals  $\tau_{\rm bs}$ when the beam-splitter unitaries act within the $|0\rangle - |1\rangle$ subspace.

\begin{figure}[htbp]
	\centering
\includegraphics[width=1\linewidth]{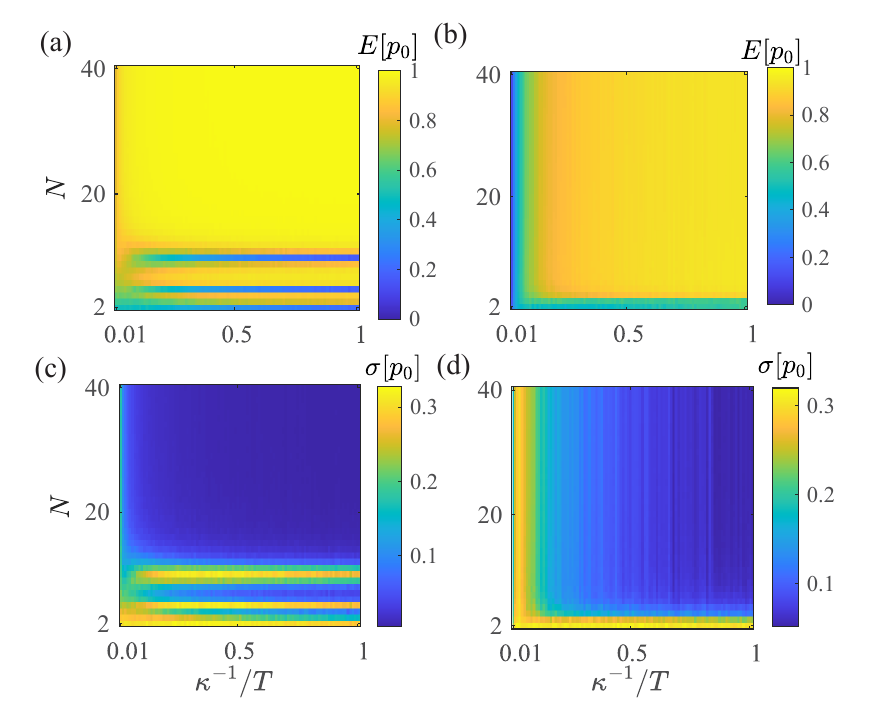}
  \caption{Panels (a) and (b) present the average of $p_0$, while panels (c) and (d) show the corresponding standard deviations. Panels (a) and (c) correspond to strongly coupled noise with $\delta \theta_p = \pm \pi/250$,
  	whereas panels (b) and (d) correspond to relatively weaker noise with $\delta \theta_p = \pm \pi/1000$.} 
 
\label{fig_noise2d}
\end{figure}

We consider the evolution of our detector qutrit under such noise (see Fig.~\ref{fig_noise}) as per the cIFM and pIFM protocols. When the number of noise samples, $\mathcal{P}$ (as described in Eq.~\ref{eq:bj}) in a pulse is much larger than $N$, the cIFM and pIFM protocols give rise to similar results. However, when $\mathcal{P} \approx N$, the cIFM and pIFM protocols can lead to quite different results. In this section, we take $\mathcal{P} >> N$, and present only the cIFM protocol to avoid any confusion. Fig.~\ref{fig_noise2d} shows the mean (E$[p_0]$) and standard deviations ($\sigma[p_0]$) of the ground state population ($p_0$) from $500$ realizations of the cIFM simulation with $T$ fixed at $10$ $\mu$s and a noise sampling rate of $10^9$ samples per second. Panels (a,c) of Fig.~\ref{fig_noise2d} correspond to effective angle $\delta \theta_p = \pm \pi/\mathcal{P}_{\rm (min)}$ and panels (b,d) correspond to $\delta \theta_p = \pm \pi/(4\mathcal{P}_{\rm (min)})$, where $\mathcal{P}_{\rm (min)} = 250$ is the number of noise samples in $\tau_{\rm B}$, corresponding to the largest $N (= 40)$ in the given range. These values of $\delta \theta_p$'s for the left and right panels of Fig.~\ref{fig_noise2d} are kept fixed throughout the simulation. Therefore, for a given $N$, the $j$th effective noise pulse angle $\theta_j$ can assume values in the range 
$[-\delta \theta_p \mathcal{P}, \delta \theta_p \mathcal{P}]$,
with discrete steps of $2\delta \theta_p$. For $N = 40$, the extreme values 
of $\theta_j$ are $\pm \pi$ for the left panel and $\pm \pi/4$ for the right panel of Fig.~\ref{fig_noise2d}, while for $N=2$, extreme $\theta_j$ values can be up to $\pm 20 \pi$ and $\pm 5\pi$, respectively. Interestingly, for a given $N$, it is very likely for $\theta_j$'s to assume the extreme values, which can also lead to certain anomalies -- as explained later in this section.

As shown in Fig.~\ref{fig_noise2d}(a), for strongly coupled noise, the mean of the marker population E$[p_0]$ swiftly approaches $1$ for small $N$ and is almost independent of $\kappa^{-1}$, with negligibly small standard deviations (Fig.~\ref{fig_noise2d}(c)), depicting a highly efficient noise detection. Despite its high efficiency, the cIFM protocol also leads to systematic anomalies, which occur due to the fact that the cIFM and pIFM protocols are transparent to values of $\theta_j$ which are an integral multiple of $4\pi$ \cite{dogra-ncomm-2022}. These anomalies manifest as horizontal lines where E$[p_0]$ almost vanishes and $\sigma[p_0]$ values are exceptionally high in Fig.~\ref{fig_noise2d}(a,c). Such situations arise for the values of $N$ that result in integral values of the ratio: $\delta \theta_p \mathcal{P}/(4\pi)$. For Fig.~\ref{fig_noise2d}(a,c), this ratio simplifies to $10/N$, leading to anomalies at $N = 1,2,5,10$. For the simulations in Fig.~\ref{fig_noise2d}(b,d), we need %$ r_i = 2.5/N$ 
$2.5/N$ to be integral, which is never satisfied, leading to no such anomalies.

 Therefore, in the case of a fast-transiting and strongly coupled noise (Fig.~\ref{fig_noise2d}(a,c)), the cIFM or pIFM protocols are quite efficient in confirming its presence, even for small values of $N$. On the other hand, if this noise, with $\theta = \pi$, interacts with a qubit on resonance, 
the qubit will typically not detect anything at all, as the noise is very likely to sum up to zero.

For relatively weakly coupled noise (Fig.~\ref{fig_noise2d}(b,d)), the marker population $p_0$ assumes higher values as $\kappa^{-1}$ increases, reflecting the Poisson point process with a smaller mean and consequently a lower switching frequency of the noise. These weaker noises also swiftly saturate the $p_0$ values for $\kappa^{-1} \geq T/5$  and $N \geq 5$. Above a certain threshold of $\kappa^{-1}$, $p_0$ shows almost no dependency on the correlation time. 
Therefore, with optimal values of $N$ and $\kappa^{-1}$, cIFM-based protocols can efficiently detect noise.

\section{Noise correlations \label{noise_correlations}}

The correlations present in the noise can often be used to reveal the underlying mechanism responsible for the fluctuations. In this section, we first show how a qubit detector can be used to measure the full counting statistics of the noise. Then, we demonstrate that 
in the case of a lower noise-sampling rate such that there is only one noise sample per $\tau_{\rm B}$ duration, correlations of random binary processes lead to different marker populations in the qutrit detectors.

\subsection{Full counting statistics with a qubit detector}

The problem of extracting the correlations of a random event is especially relevant in mesoscopic physics, where the challenge of measuring the statistics of electronic transport in nanoelectronic devices has led to the so-called problem of full counting statistics \cite{Belzig2005}. In full counting statistics, we are interested in the probability distribution $P(m,T)$ of events $m$ in a given time interval $T$. The complete information about correlations is encapsulated in the generating function, defined as the Fourier transform of $P(m,T)$, which allows us to calculate arbitrarily high-order cumulants associated with $P(m,T)$. The compact variable of this transform, called the counting field, can  be understood as a variable coupling between the noise and a detector. For example, in proposals that use a qubit to characterize the statistics of electrons transmitted through a quantum point contact, the counting field is the coupling between the current generated by the electrons and the $\sigma_z$ operator of the qubit \cite{Lebedev2016}. We now consider the qubit detector as described above and ask the question: what is the signature of higher-order cumulants of amplitude noise in the measured signal?  

A straightforward realization of these events in our qubit-detector setup is to take a series of pulses $\theta_j \in\{ 0, \theta \}$ distributed in accordance with the probability $P(m,T)$,
and introduce a scaling factor $\lambda$ which can serve as the counting field. In practice, $\lambda$ can be realized simply by introducing a variable attenuator between the noise to be detected and the qubit. We consider the time interval of a full  sequence $T = \tau_{\rm bs}(N+1) + N\tau_{\rm B}$ and count how many times $m$ we had a non-zero $\theta$, with the total angle accumulated being $\theta_{T} = \int_{0}^{T} \Omega (t) dt = \sum_{i=1}^{N} \theta_{j} = m\theta$. 

The generating function can be defined as
\begin{equation}
	\Lambda_{\theta} (\lambda) = \langle e^{i \lambda \theta_{T}} \rangle = \sum_{m} P(m,T) e^{i m \lambda \theta}, \label{eq:defi}
\end{equation}
from which the $k$th order moments $\langle \theta_{T}^k \rangle$ of the total angle can be obtained by
\begin{equation} 
\langle \theta_{T}^k \rangle =\langle m^k \rangle \theta^k =  (-i)^k \lim_{\lambda \rightarrow 0} \partial^k_{\lambda} \Lambda_{\theta }(\lambda ) . \label{eq:mom} 
\end{equation}

A qubit detector would then have a probability $p_{\rm e}(m\lambda \theta)$ of ending up in the marker state $|{\rm e}\rangle$ if there are $m$ events, leading to an overall average probability $E[p_{\rm e}](\lambda\theta ) = \sum_{m} P(m,T) p_{\rm e} (m\lambda \theta )$ for the entire ensemble. Let us assume, for simplicity, that the coupling of the  $B$ pulses is via the vector $\hat{\bf n}_j =(0,1)$ for all $j$'s, in other words, $\varphi_j = - \pi/2$ [see Eq. (\ref{eq:bpls})].
If the qubit starts in the state $|{\rm g}\rangle$, we obtain that the probability of having $|{\rm e}\rangle$
 is
\begin{equation}
	p_{\rm e} (m\lambda \theta ) = \frac{1}{2} \left[1 - \cos (m \lambda \theta )\right],
\end{equation}
therefore,
\begin{equation}
	E[p_{\rm e}] (\lambda\theta ) = \frac{1}{2} \sum_{m} P(m,T) [1 - \cos (m \lambda \theta )] = \frac{1}{2}
	\left[ 1 - \Re( \Lambda_{\theta}  (\lambda))\right].	\label{eq:real}
\end{equation}
Similarly, starting with the state 
 $(1/\sqrt{2})[|{\rm g}\rangle + |{\rm e}\rangle ]$, we obtain 
\begin{equation}
	p_{\rm e} (m\lambda \theta ) = \frac{1}{2} \left[ 1 + \sin (m \lambda\theta ) \right],
\end{equation}
and therefore,
\begin{equation}
	E[p_{\rm e}] (\lambda\theta ) = \frac{1}{2}\sum_{m} P(m,T) [1 + \sin (m \lambda \theta )] = \frac{1}{2}
	\left[ 1 + \Im( \Lambda_{\theta}  (\lambda))\right]	. \label{eq:imag}
\end{equation}
This means that we can directly obtain both the real and imaginary part of the moment generating function by measuring the average probability $E[p_{\rm e}]$ with two different initial conditions. We can then repeat this for various values of $\lambda$ (which can be varied by using an appropriate attenuator), obtain an approximate functional dependence $\Lambda_{\theta} (\lambda )$, and extract the moments using Eq.~(\ref{eq:mom}).

Full counting statistics offers a different perspective on characterizing noise than the usual analysis of correlations, by counting events in a time interval. The two perspectives are, of course, connected to one another, although the relationship may not always be simple \cite{Belzig2005}. For example, in our case, the second-order moments can be connected to the zero-frequency power spectral density
\begin{align}
	\langle \theta_{T}^2 \rangle & =\int_{0}^{T} \int_{0}^{T}  \langle \Omega (t_1)\Omega (t_2) \rangle dt_{1}dt_{2} \\
	 & = T \int_{-\infty}^{\infty} \langle \Omega (0) \Omega (\tau ) \rangle d\tau  =  T S_{\Omega} (f=0),
\end{align}
where the second row is obtained by a change of variables $T = (t_1+ t_2)/2$, $\tau = t_2 - t_1$, with $T$ assumed to be large.

For example, consider the Poisson distribution $P(m,T) = (\kappa T)^{m}\exp (-\kappa T)/m!$. The generating function is obtained from the definition Eq. (\ref{eq:defi}) as

	\begin{equation}
		\Lambda_{\theta} (\lambda ) = \exp \left[\kappa T (e^{i\lambda \theta} -1) \right].
	\end{equation}
This generating function can be obtained experimentally by following the protocol described above and using Eqs. (\ref{eq:real}, \ref{eq:imag}).
In particular, from Eq. (\ref{eq:mom}), we find $\langle \theta_{T}^2 \rangle = \kappa T \theta^2 + (\kappa T \theta)^2$, demonstrating that the variance of a Poisson distribution equals its mean, as expected.
This approach allows us to extract the rate $\kappa$ and also characterize the zero-frequency power spectral density of the underlying noise in $\Omega$.

\subsection{Signatures of correlations in cIFM }

In the previous subsection, we have seen that the response of the qubit does not depend on how the $m$ occurrences of $\theta$ pulses are distributed in the time interval $T$: they would simply sum up to $m\theta$ and the response would be a sine or cosine of $m\theta$. This is not the case for cIFM,
which is sensitive to how these events are correlated.
 To illustrate this, consider a uniform distribution of $\theta$ values over the $N$ $\tau_{\rm B}$ durations. In this case, for cIFM, we have $S(\phi_{N}) [B(\theta ) S(\phi_{N})]^N |0\rangle$ as the final state and can utilize the results for the probability amplitudes from Ref.~\cite{jake-prr-2023}. Let us now consider he opposite situation: we concentrate all the driving power in one single interaction with the qutrit, occurring after the $n$th
application of $S(\phi_{N})$ ($0<n<N$). We obtain the final state $[S(\phi_{N})]^{N+1-n}B(N \theta)[S(\phi_{N})]^{n}|0\rangle = c_{0}|0\rangle + c_{1}|1\rangle + c_{2}|2\rangle $, with probability amplitudes
\begin{eqnarray}
c_{0} &=& \sin (n\phi_{N}) \sin^{2} \frac{N\theta}{4}, \\	
c_{1} &=& \cos^2 \frac{N\theta}{4} + \cos(n \phi_{N})\sin^2 \frac{N\theta}{4}, \\
c_{2} &=& \sin \frac{N\theta}{2} \sin \frac{n\phi_{N}}{2}.
\end{eqnarray}

In comparison with the uniform case, the differences are significant. For example, increasing $N$ at fixed $m$ does not suppress the coefficient $c_1$ as in the uniform case. In fact, at large $N$ we would get $c_{0} \simeq 0$, $c_{1} \simeq 1$, $c_{2}\simeq 0$, so the detection signal produced is the same as for the case when no pulse is present. In other words, the detector completely misses the extremely strong $N\theta$ pulse.

Now consider the more realistic case of $N=4$, which has four $B$ pulse slots and five
beam-splitter pulses. As per the cIFM protocol, the unitary evolution can be explicitly represented as: $S(\pi/5) B(\theta_4) S(\pi/5) B(\theta_3)S(\pi/5) B(\theta_2)S(\pi/5) B(\theta_1)S(\pi/5)$, see Fig.~\ref{detection_schemes}(b). Let us fix two of these $B$ pulse angles at $\theta$ and set the remaining two to zero. There are six possible combinations, shown in the second column of Table~\ref{table1}. Corresponding to each of these configurations, the marker population ($p_0$) values for cIFM and pIFM are specified. Clearly, these values differ markedly across configurations, with significant differences for cIFM and relatively smaller differences for pIFM. This signifies the role of correlations between the pulses in the cIFM and pIFM protocols. In the lower part of Table~\ref{table1}, we also consider another set of combinations of $\theta$ values, where two of these values are $\pi$ and the other two are $-\pi$. Again, in cIFM, we observe clear differences in $p_0$ values for different configurations, while pIFM is insensitive to the correlations in this case.

This feature means that under certain conditions, cIFM can distinguish between clustered noises and other arbitrarily-correlated noises. 
To exploit this feature of cIFM, one must consider a lower sampling rate, such that there is only one noise sample ($\mathcal{P}=1$) in one whole $\tau_{\rm B}$ duration. Otherwise, noise amplitudes in the given pulse duration get averaged, leading to $\mathcal{P}+1$ possibilities of $\theta_j$ values, and hence the clustering patterns of the original binary noise waveform will be lost.
To illustrate this, we simulate binary noise with amplitudes $ \theta_j = \pm \pi$, assuming that each noise amplitude stays constant within a $B$ pulse duration. For an arbitrary value of $N$, we generate $m$ events using the Poisson point process as described in Sec.~\ref{binary_noise}, with a noise sampling rate of $N\tau_{\rm B}/T$ corresponding to different values of 
 {$\kappa^{-1} \in [T/10, T]$
 and observe the ground state populations for different values of $\kappa^{-1}$ and $N$. The results are shown in Fig.~\ref{fig_noise3}, where the panel on the right (b) presents an example of the binary noise for $N=10$ at 
 $\kappa^{-1} = T/10$ (first column, labeled as $\kappa_{\rm max}$) and at $\kappa^{-1} = 10 \tau_B = T$ (third column, labeled as $\kappa_{\rm min}$), while the noise for an arbitrary $\kappa^{-1}$ is shown in the second column. 
 The mean value of the Poisson distribution ($\kappa T$) is quite different in the two extreme situations, as reflected in the nature of these noises. For larger $\kappa^{-1}$, it is more likely to have less frequent switching of the noise amplitude, leading to more clustering of the noise pulses. Fig.~\ref{fig_noise3}(a) presents the mean value of $p_0$ obtained from several repetitions of the cIFM protocol as a function of correlation time $\kappa^{-1}$. Different curves correspond to different values of $N$, as specified in the plot legends. As we move from left to right, the same amplitude values of the noise are more likely to be clustered together, leading to higher $p_0$ values. Also note that as $N$ increases, the detection becomes more insensitive to the underlying correlations of the noise. 
 
 \begin{figure}
	\centering
\includegraphics[width=1\linewidth]{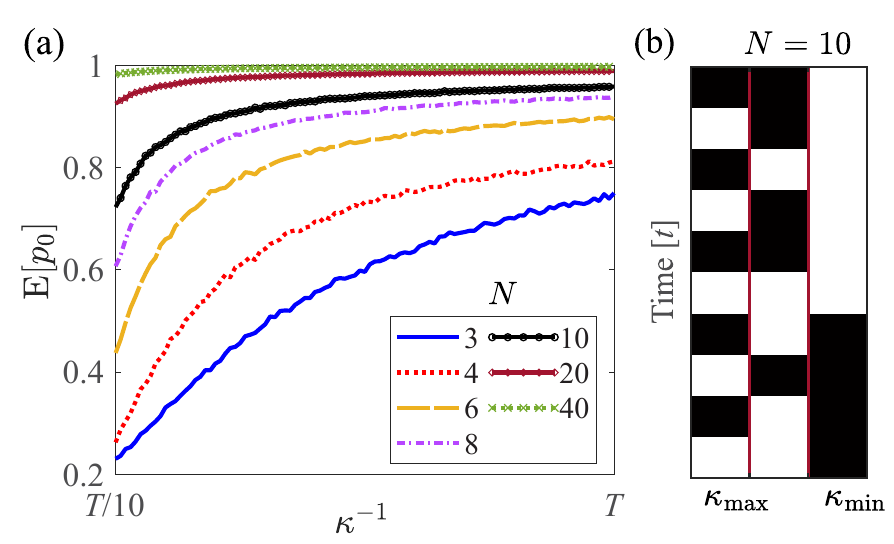}
  \caption{(a) Average of $p_0$ over $2000$ realizations as a function of $\kappa^{-1}$, with different curves corresponding to various values of $N$.
  	 (b) Illustration of a binary random process for $N=10$ for three different values of the switching frequency $\kappa$, where the trace on the extreme right demonstrates clustering of the pulses.}
\label{fig_noise3}
\end{figure}

In contrast, the pIFM protocol is not sensitive to this particular type of correlation, as we could anticipate from the previous analysis in Table~\ref{table1}. In the example above, the noise can also be regarded as binary noise with constant amplitude $\theta_{j} = \mathrm{const.}$ and flipping phase $\varphi_j = \pm \pi$. Since the pIFM protocol is not sensitive to phase information, noises with different correlations cannot be distinguished. In the pIFM protocol, this noise does not lead to a distinct signal from that of a train of $\pi$ pulses, which gives rise to a constant detector qutrit ground state population, $p_0 = \cos^{2(N+1)}(\pi/(2(N+1)))$ for an arbitrary $N$~\cite{dogra-ncomm-2022, jake-prr-2023}. We have also verified this result numerically, an example of which can be seen in rows numbered 7--10 of Table~\ref{table1}.

\begin{table}[H]
	\setlength\extrarowheight{3pt}   
	\begin{center}
		\begin{tabular}{|c|c|c|c|}
			\hline
			\hspace{5mm} & \textbf{Configuration} & \hspace{3mm}\textbf{cIFM} \hspace{3mm}& \hspace{3mm} \textbf{pIFM} \hspace{3mm}\\
			\textbf{} & $\theta_1$, $\theta_2$, $\theta_3$, $\theta_4$ & $p_0$ & $p_0$ \\
			\hline
			\textbf{$1$} & $\pi,\,\pi,\,0,\,0$ & $0.611$ & $0.283$ \\[5pt]
			\hline
			\textbf{$2$} & $\pi,\,0,\,\pi,\,0$ & $0.646$ & $0.387$ \\[5pt]
			\hline
			\textbf{$3$} & $\pi,\,0,\,0,\,\pi$ & $0.393$ & $0.283$ \\[5pt]
			\hline
			\textbf{$4$} & $0,\,\pi,\,\pi,\,0$ & $0.937$ & $0.387$ \\[5pt]
			\hline
			\textbf{$5$} & $0,\,\pi,\,0,\,\pi$ & $0.646$ & $0.387$ \\[5pt]
			\hline
			\textbf{$6$} & $0,\,0,\,\pi,\,\pi$ & $0.611$ & $0.283$ \\[5pt]
			\hline 
			\hline
			\textbf{$7$} & $\pi,\,\pi,\,-\pi,\,-\pi$ & $0.599$ & $0.605$ \\[5pt]
			\hline
			\textbf{$8$} & $\pi,\,-\pi,\,\pi,\,-\pi$ & $0.183$ & $0.605$ \\[5pt]
			\hline
			\textbf{$9$} & $\pi,\,-\pi,\,-\pi,\,\pi$ & $0.361$ & $0.605$ \\[5pt]
			\hline
			\textbf{$10$} & $-\pi,\,\pi,\,\pi,\,-\pi$ & $0.361$ & $0.605$ \\[5pt]
			\hline
			\textbf{$11$} & $-\pi,\,\pi,\,-\pi,\,\pi$ & $0.183$ & $0.605$ \\[5pt]
			\hline
			\textbf{$12$} & $-\pi,\,-\pi,\,\pi,\,\pi$ & $0.599$ & $0.605$ \\[5pt]
			\hline
		\end{tabular}
		\caption{Marker populations $p_0$ resulting from cIFM and pIFM protocols for $N=4$, with the qutrit initialized in state $|0\rangle$.}
		\label{table1}	
	\end{center}	
\end{table}

\section{Applications}
\label{appl}

The cIFM noise-sensing protocol can be adapted to a variety of experimental platforms where a controllable three-level system exists. Below, we give two such examples: flux qutrits and Rydberg systems. Another immediate implementation could be in trapped ions, which has already been investigated in the context of realizing interaction-free CNOT gates \cite{Azuma2003, Pavicic2007}.

The cIFM protocol has already been implemented in a transmon qubit \cite{dogra-ncomm-2022}, and noise detection in this setup would be straightforward. Instead, we discuss here a different superconducting qubit -- a flux qubit -- which, by virtue of its large anharmonicity, would allow us to access noise around higher frequencies and in larger bandwidths than a transmon ($\sim$ 300 MHz) \cite{Yan2016, Koch2007}. Due to this sizeable anharmonicity, a flux qutrit will have a reduced coupling of the noise into its lower transition, making it principally a more suitable candidate for implementing cIFM noise detection.

The level separations can be adjusted by changing the external magnetic flux applied to the qutrit loop. When the reduced magnetic flux $\Phi_{\rm ext}/\Phi_{0}$ is a half-integer value, the potential energy is symmetric. Here, $\Phi_{\rm ext}$  is the external magnetic flux threading the superconducting loop, and $\Phi_{0} = h/2e$ is the flux quantum. At these so-called \textit{sweet spots} in the reduced magnetic flux, where the potential energy is symmetric, the energy levels of the qubit are less sensitive to small variations in the external magnetic flux. This is because the first derivative of the energy levels with respect to the flux bias is zero. As a result, the qubit's transition frequency is less affected by flux noise, which typically manifests as low-frequency ($1/f$) noise \cite{Yoshihara_2006}. Since flux noise can cause fluctuations in the qubit's transition frequency, leading to dephasing, the qubit is also less sensitive to flux noise at the sweet spots, resulting in longer dephasing times \cite{Ithier_2005}. This means the qubit can maintain coherent superpositions for a longer duration, improving the performance of quantum operations.

For a three-junction flux qubit, the typical frequency range of the energy level splitting at the symmetry point, i.e., sweet spot, is generally within the range of a few GHz. Specifically, the transition frequency typically falls in the range of approximately 5 to 10 GHz \cite{Yan2016}.

To calibrate this system, external noise around the qubit frequency can be generated artificially and coupled into the qubit. This type of noise injection has already been utilized for characterizing dephasing noise using transmon qubits \cite{Pappas2022}. Alternatively, if the flux qutrit is coupled to a resonator for readout, noise can be injected via the resonator \cite{Yan2016}.
Readout can be performed by inductively coupling the flux qutrit to a dc-superconducting quantum interference device (SQUID) and measuring the switching currents \cite{Lupascu_2005}, or alternatively, by using dispersive readout via a resonator \cite{Bianchetti_2010}.

\begin{figure}
	\centering
	\includegraphics[width=1\linewidth]{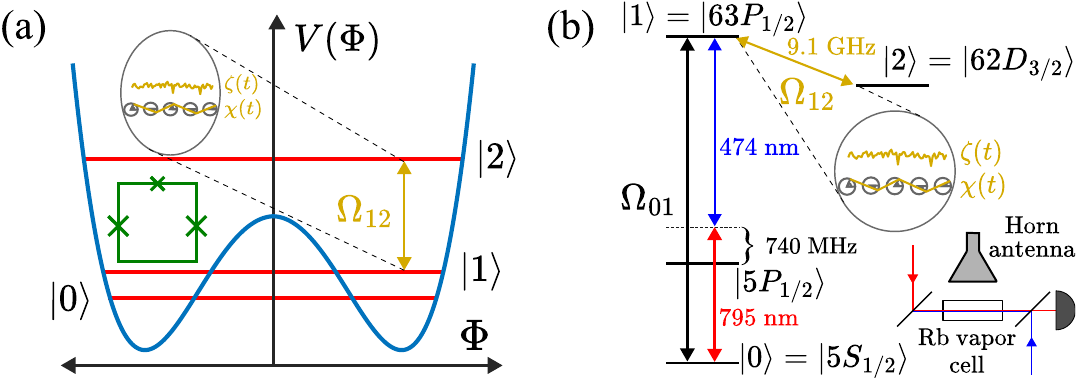}
	\caption{(a) A schematic showing how cIFM noise detection could be adopted for a flux qutrit, where the pronounced anharmonicity is exploited to detect noise resonant with the $\vert1\rangle - \vert2\rangle$ transition. The inset on the left side of the potential depicts a schematic of a flux qutrit with three Josephson junctions, while the inset extending from the right illustrates  amplitude $\zeta(t)$ and phase $\chi(t)$ noise. (b) A schematic demonstrating the adaptation of our noise detection protocol to Rydberg atoms, specifically $^{87}$Rb, driven by a two-photon process detuned by $750$ MHz from the intermediate state $\vert 5P_{1/2}\rangle$. The atoms can be placed in a vapor cell or optical-tweezer trap, with an avalanche photodiode used to monitor probe laser beam transmission. Microwave noise can be coupled in using a dipole or horn antenna.
	}
	\label{fig_apps}
\end{figure}

Our cIFM noise detection protocol can also be adapted for sensing with Rydberg atoms. These atoms are in highly excited states, which consequently makes them extremely sensitive to microwave fields; indeed, the dipole moment between nearby states scales as $\sim n^2$ and the polarizability scales as $\sim n^7$ \cite{Browaeys_2016}. The transition energies between adjacent Rydberg states span a broad spectrum, ranging from MHz to THz, allowing us to access high frequencies where conventional off-the-shelf electronics are not available \cite{He2021}.

A schematic for implementing cIFM is shown in Fig. \ref{fig_apps} (b). The atoms can be placed either in a vapor cell or loaded into an optical-tweezer trap. We identify $|0\rangle = |5S_{1/2}\rangle$, $|1\rangle=|63P_{1/2}\rangle$, and $|2\rangle =|62D_{3/2}\rangle$. Due to their frequency difference being in the ultraviolet range, it is practically convenient to couple states $|0\rangle$ and $|1\rangle$ using a two-photon process driven by standard optical lasers with wavelengths 795 nm and 474 nm, with $|5P_{1/2}, F=2, m_{F}=2\rangle$ as the intermediate state (single-photon detuning of $740$ MHz). In this setup, typical values for the effective two-photon Rabi frequency $\Omega_{01}/(2\pi )$ range from 500 kHz to 5 MHz \cite{Browaeys2010,Browaeys2013}, while $\Omega_{12}/(2\pi)$ is approximately $5-7$ MHz \cite{Browaeys2015}. The time $T$ of our protocol is limited by the finite lifetime of the Rydberg states and off-resonant excitations on $|5P_{1/2}, F=2, m_{F}=2\rangle$; in practice, $T\sim 4$ $\mu$s allows for the maximum value of $\theta = \pi$, enabling the realization of 
around $N\sim 40$ pulses, each with a duration of about 100 ns \cite{Browaeys2015}.
The readout scheme is based on monitoring the optical transmission through the atomic gas in vapor cells \cite{He2021,Kumar17,Shaffer2012} or, in the case of optical tweezers, on the fact that Rydberg states are not trapped; therefore, the signal in fluorescence measurements disappears unless the atoms are in the ground state \cite {Browaeys2010,Browaeys2013,Browaeys2015}. Tunability in terms of the 1--2 frequency can be achieved in this system either by resorting to the Zeeman effect or in discrete steps using different Rydberg transitions.

\section{Conclusions \label{conclusions}}

Characterizing noise at certain frequencies is essential for the development of quantum technologies. By using interaction-free measurements implemented with a qutrit, we demonstrate the ability to sense low-intensity noise and observe features that depend on correlations. This is compared with the case of a single detector qubit, the simplest example of an absorption detector, where noise creates an excitation that can be subsequently observed.

In a qubit-based detector, noise detection characterized by small correlation times results in the system being driven towards a maximally mixed state. This state corresponds to a situation where the probabilities of finding the qubit in either of its basis states ($|{\rm g}\rangle$ or $|{\rm e}\rangle$) are equal, leading to a marker population of 0.5. In contrast, qutrits, having an extra degree of freedom, allow for a more sophisticated noise detection protocol where detection does not result in any excitations. We find that for a variety of noise types, these interaction-free measurements are much more effective.

The application of cIFM and pIFM protocols leads to high-purity states with marker populations approaching 1, while the absence of noise is characterized by $p_0 = 0$. The efficiency of the cIFM and pIFM protocols increases with $N$, as evident from the increasing mean values and almost diminishing variance in marker populations, indicating that only a few repetitions of the detection protocol are sufficient to detect the presence of noise.

While a qubit detector can measure the full counting statistics of noise events in a given time interval, the cIFM detector is also sensitive to how these events are correlated and is effective at distinguishing clustered noises and other arbitrarily correlated noises.

In essence, cIFM-based protocols are more robust and versatile in efficiently detecting resonant noise. Our results are general and applicable to any experimental platform where interaction-free measurements can be implemented.

\acknowledgments

We acknowledge financial support from Business Finland QuTI (decision 41419/31/2020) as well as from the Research Council of Finland Centre of Excellence program (Project No. 352925).

\appendix
\section{Noise characteristics}

For completeness, we give a brief presentation of the notations and concepts related to noise utilized in this work. In general, for a random time-dependent variable $X(t)$, we define the double-sided power spectral density at a frequency $f$ as 
\begin{equation}
	S_{X}(f) = \int_{-\infty}^{\infty} R_{XX}(\tau) e^{- i 2 \pi f \tau } d \tau ,\label{eq:WK}
\end{equation}
where $R_{XX}(\tau ) = \mathbb{E}[X(t) X(t + \tau)]$ is the autocorrelation function.
Alternatively, this Fourier-transform connection between autocorrelation and power spectral density can be introduced as the result of the Wiener-Khinchin theorem. For ergodic processes, the ensemble average is identical to the 
time-average in a window defined from $-W/2$ to $W/2$, therefore $R_{XX}(\tau) = \lim_{W\rightarrow \infty}\frac{1}{W}\int_{-W/2}^{W/2}X(t)X(t+\tau) dt$.
Moreover, the average power $\overline{X^2}$ can be obtained as $\overline{X(t)^2} = R_{XX}(0) = \int_{-\infty}^{\infty} S_{X}(f) df$. If $X(t)$ is a real random variable, then it follows directly from the definitions that both $R_{XX}$ and $S_{X}(f)$ are even and real-valued. 
The power spectral density can more conveniently be obtained from the windowed Fourier transform 
\begin{equation}
	X_{W} (f) = \int_{-W/2}^{W/2} dt X(t) e^{-i 2 \pi ft},
\end{equation}
as 
\begin{equation}
S_{X}(f) = \lim_{W\rightarrow \infty} \frac{1}{W} |X_{W}(f)|^2 .
\end{equation}

From here, we see that $S_{X}(f)$ is also positive. Since the frequency $f$ is positive in real experiments, it is convenient to introduce the single-sideband power spectral density defined by $\mathbb{S}_{X} (f) = 2 S_{X}(f)$ if $f>0$, and zero otherwise (see, e.g., Ref.~\onlinecite{Cattaneo2021} for more details).

For example, consider the Poissonian process introduced in Sec. \ref{binary_noise}. In a time interval $\tau$, the probability of $m$ events is $P(m,\tau) = (1/m!)(\kappa \tau)^m e^{- \kappa\tau}$. Here, the events are defined by the variable $X$ taking one of the two discrete values $x$ or $-x$. To calculate the autocorrelation function $R_{XX}(\tau)$, we consider a time interval $\tau$ and separate the probabilities corresponding to an even number of switches (producing an $x^2$ value in the autocorrelation) and the probabilities corresponding to an odd number of switches (producing a $-x^2$ value). Thus, $R_{XX}(\tau) = x^2 (P (0,|\tau|) + P (2,|\tau|) + ...) - x^2 (P (1,|\tau|) + P (3,|\tau|) + ...)$. Here, the modulus appears because, in general, $\tau$ can be negative.
This yields the double-sided exponential:

\begin{equation}
R_{XX}(\tau) = 	x^2 \exp (-2 \kappa |\tau|).
\end{equation}

The clustering effect can be observed by defining the normalized correlation $R_{XX}(\tau)/R_{XX}(0) = 
 \exp (-2 \kappa |\tau|) < 1$. 
We can calculate the spectrum $S_{X}(f)$ by explicitly calculating the integral Eq.~(\ref{eq:WK}). 
Note that the result is real because we can write  $\exp (- i 2 \pi f \tau ) = \cos (2 \pi f \tau ) - i \sin (2 \pi f \tau)$, and the integral containing the sine vanishes due to the $\tau \rightarrow - \tau$ antisymmetry. From the remaining cosine part, we obtain the Lorentzian
\begin{equation}
	S_{X}(f) = \frac{x^2 \kappa}{\kappa^2 + \pi^2 f^2}.
\end{equation}		
In the case of a generic drive with nominal frequency $\omega_{0}$ and Rabi coupling $\Omega (t)$, we can write  
$\Omega(t)\cos (\omega_{0}t + \chi (t))$, where $\chi (t)$ is the phase noise, and $\Omega(t) = \Omega+ \zeta(t)$, with amplitude noise $\zeta (t)$.  The Rabi coupling is also noisy, with an associated double-sided spectral density
\begin{equation}
	S_{\Omega}(f) = \lim_{W\rightarrow \infty}\frac{1}{W} |\Omega_{W}(f)|^2 ~~~~ (\mathrm{Hz}).
\end{equation}
The double-sided spectral density of the phase noise is
\begin{equation}
	S_{\chi}(f) = \lim_{W\rightarrow \infty}\frac{1}{W} |\chi_{W}(f)|^2 ~~~~ (\mathrm{rad}^2/\mathrm{Hz}),
\end{equation}
which is typically expressed in decibel units $\mathrm{dB_{c}}/\mathrm{Hz}$ by applying the logarithmic scaling  $10 \log_{10}$. Here, the subscript $c$ denotes the carrier.

It is also convenient to introduce the fractional frequency noise, defined via the random variable $y (t) = \Delta \omega (t)/\omega_{0}$, where $\Delta \omega (t) = \omega (t) - \omega_{0}$ and $\omega (t) = \frac{d}{dt}[\omega_{0}t +\chi (t)] = \omega_{0} + \dot{\chi}(t)$; the power spectral density of the fractional frequency noise is given by
\begin{equation}
	S_{y} (f) = \frac{(2 \pi f)^2}{\omega_{0}^2} S_{\chi}(f) .
\end{equation} 
Different power laws as a function of frequency $f$ can be obtained depending on the mechanism, with various types of noise dominating at low, intermediate, or high frequencies. The primary types encountered in oscillators, listed in increasing frequency scaling of the PSD, include: a random walk of frequency [$S_{y} (f) \sim 1/f^{2}$, $S_{\chi} (f) \sim 1/f^{4}$]; 
frequency flicker [$S_{y} (f) \sim 1/f$, $S_{\chi} (f) \sim 1/f^{3}$]; random walk of phase (brown phase noise) or white noise of frequency [$S_{y} (f) \sim {\rm const.}$, $S_{\chi} (f) \sim 1/f^{2}$]; phase flicker or pink phase noise [$S_{y} (f) \sim  f$, $S_{\chi} (f) \sim 1/f$]; and white phase noise [$S_{y} (f) \sim  f^2$, $S_{\chi} (f) \sim {\rm const.}$].

Moreover, noise is often characterized according to the frequency scaling of its power spectral density (PSD), known as its \textit{color}. White noise, characterized by a constant power spectral density and often simulated using a random number generator, has equal power at all frequencies. 
  Filtered white noise is referred to as colored or \textit{correlated} noise \cite{Smith2011}, resulting from the convolution of white noise with an impulse response. Colored noise, in general, can be created by applying a Fourier filter of specific power.
  
		Brown noise, the integral of white noise, exhibits an amplitude response proportional to $1/f$ and is typically generated by low-pass filtering white noise. Its power spectral density decreases by  $20\log_{10}(0.5) = -6.02$ dB per octave or $-20$ dB/decade. Pink noise, often referred to as $1/f$ noise due to its PSD proportionality, cannot be perfectly obtained by filtering white noise because the filter's amplitude response must scale as $f^{-1/2}$ \cite{Smith2011}. However, it can be approximated by filtering uniformly distributed random numbers through a finite impulse response (FIR) filter with a $1/f$ passband. Pink noise exhibits a power spectral density decreasing by approximately $10\log_{10}(0.5)=-3.01$ dB per octave or $-10$ dB per decade (see Fig.~\ref{fig_pink_noise}). 
		
		Blue noise, which has a power spectral density that scales linearly with frequency, can be efficiently generated using Poisson disk sampling \cite{Bridson2007}. It exhibits an approximate decrease of $3.01$ dB per octave (10 dB per decade). Purple noise, the derivative of white noise, has a power spectral density that decreases by approximately $6.02$ dB per octave (20 dB per decade) and can be generated by combining blue noise and brown noise or using a band-stop filter.

\begin{figure}
	\centering
	\includegraphics[width=0.97\linewidth]{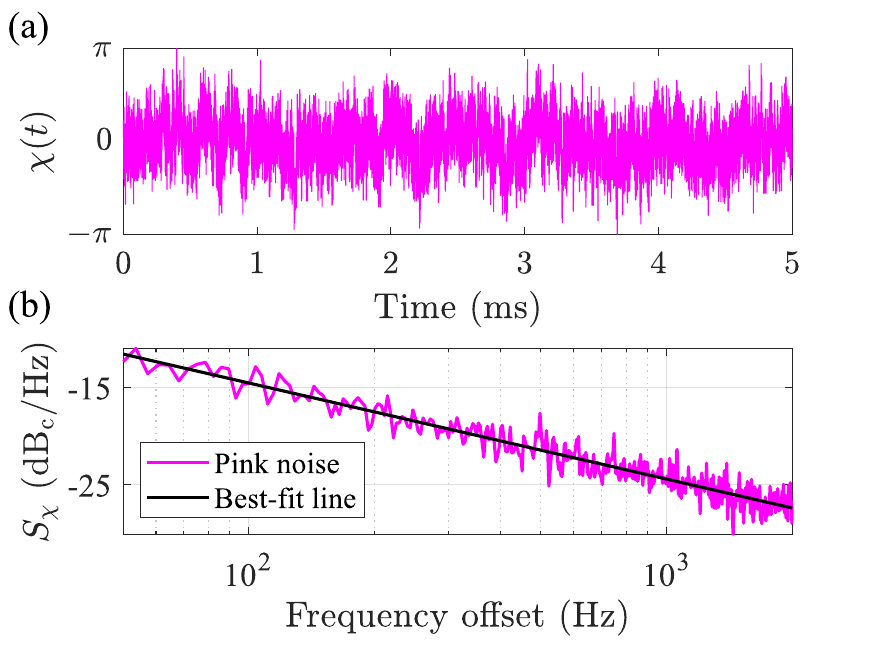}
	\caption{A pink phase (phase flicker) noise $\chi(t)$ with $5 \times 10^{4}$ samples. Panel (a) shows the noise in the time domain, and (b) displays its power spectral density (PSD) with frequency offsets at $0.1$ and $1$ kHz. The PSD confirms the noise's pink nature, as indicated by the semi-log fit showing a decrease of approximately 10 $\rm dB_{c}$/decade.}
	\label{fig_pink_noise}
\end{figure}

\begin{figure}
	\centering
	\includegraphics[width=0.97\linewidth]{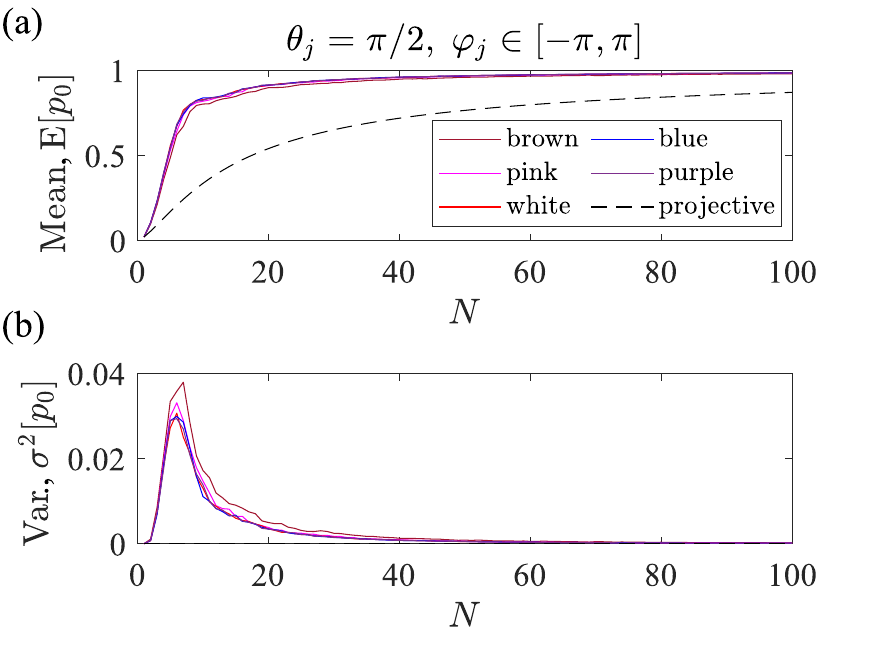}
	\caption{(a) cIFM marker probabilities (solid curves) vs $N$ at $\theta_{j} = \pi/2$ and various colors of phase noise $\varphi_{j} \in [-\pi, \pi]$ averaged over $5 \times 10^{4}$ realizations. Here, there is only one noise sample per $B$ pulse. Note that the pIFM marker probabilities (dashed black curve) are insensitive to phase and thus are the same regardless of the phase noise color. Panel (b) shows the corresponding variances.}
\label{fig_phase_noise_colors}
\end{figure}

Noises with different spectral exponents can be generated by considering the Poissonian process discussed in Secs. \ref{binary_noise} and  \ref{noise_correlations}, incorporating an appropriate power-law distribution of the characteristic correlation time $\tau$ \cite{Ziel1979, Dutta1981, Paladino2014}. 

In Fig.~\ref{fig_phase_noise_colors}, we compare the marker probabilities of pIFM with those of cIFM at different phase noise colors, where there is only one noise sample per $B$ pulse  ($\mathcal{P}=1$). In particular, we observe marker probabilities for brown noise [$S_{\chi} (f) \sim 1/f^{2}$],  pink noise [$S_{\chi} (f) \sim 1/f$], white noise [$S_{\chi} (f) \sim {\rm const.}$], blue noise [$S_{\chi} (f) \sim f$], and purple noise [$S_{\chi} (f) \sim f^{2}$]. From  Fig.~\ref{fig_phase_noise_colors}, it is evident that the cIFM protocol is not effective at distinguishing between the different colors of phase noise, as the mean marker probabilities E$[p_0]$ (averaged over $5 \times 10^{4}$ realizations) cluster tightly together with variances peaking at approximately the same $N$.

\section{Comparison between cIFM and qubit-based protocols}

To understand why the cIFM detector performs better than the qubit detector, especially in the case of alternating-sign binary noise, consider the case $N=2$. Here, 
$S_{2}$ is realized by a  $\phi_2 = \pi/3$ pulse and takes the form:
\begin{equation}
	S_{2} = \frac{\sqrt{3}}{2} \mathbb{I}_{01} -  \frac{i}{2}\sigma_{01}^{y} +|2\rangle \langle 2|.  \label{Eq-S2}
\end{equation}
Let us now assume the first $B$ pulse has angle $\theta >0$, while the second one has angle $-\theta$. For a qubit detector, this would result in a complete cancellation of the detection signal.
However, with cIFM, the state after applying the algorithm is: 

\begin{equation}
	\begin{split}
		\left(\frac{3\sqrt{3}}{8} -\frac{2\sqrt{3}}{8} \cos \frac{\theta}{2} - \frac{\sqrt{3}}{8}\cos^{2} \frac{\theta}{2} - \frac{1}{4}\sin^2 \frac{\theta}{2} \right)|0\rangle  \\
		+ \left( \frac{3}{8}  + \frac{2}{8} \cos \frac{\theta}{2}  + \frac{3}{8}\cos^{2} \frac{\theta}{2} 
		+ \frac{\sqrt{3}}{4} \sin^2 \frac{\theta}{2} \right)|1\rangle  \\
		 + \left(\frac{2-\sqrt{3}}{4} \sin \frac{\theta}{2} \cos \frac{\theta}{2} - \frac{\sqrt{3}}{4}\sin \frac{\theta}{2} \right)|2\rangle\;. \label{eq:N2semiclassical}
	\end{split}
\end{equation}
One clearly sees that the amplitude probability for the state $|0\rangle$ is not zero. Even for $\theta \ll 1$, we can approximate the state as: 
\begin{equation}
	-\frac{\theta^2}{16}|0\rangle + \left[1 + \frac{(2+\sqrt{3})\theta^2}{16}\right]|1\rangle + 
	\frac{\theta}{4}(1 - \sqrt{3})|2\rangle ,
\end{equation}
showing that the amplitude for $|0\rangle$ is second-order in $\theta$, but not zero.

Now, consider the general case of sampling the noise with values $\theta$ and $-\theta$ with equal probability using $N+1$ beam-splitter unitaries. If $N=2$, as above, the sampling space consists of $(+\theta, +\theta)$, $(+\theta , -\theta)$, $(-\theta , +\theta )$, and $(-\theta , -\theta )$. The alternating-sign situations occur with the same  probability as the same-sign situations, hence there is no clear advantage for cIFM. However, when $N$ gets large, the number of cases where the sum of $\theta_{j} = \pm \theta$ is $k$, i.e., $\sum_{j=1}^{N}\theta_{j} = (N-k) \theta + k (-\theta )=(N-2k)\theta$, is 
determined by the binomial coefficient $C^{N}_{k}$. Specifically, the probability distribution is binomial:

\begin{equation}
	p (k) = \frac{1}{2^N} C^{N}_{k} .
\end{equation}
In the limit of large $N$, this approximates to a normal distribution:
\begin{equation}
	p(k) = \frac{2}{\sqrt{\pi N}}e^{-2\left(k -\frac{N}{2}\right)^2/N} .
\end{equation}
The maximum of $p$ occurs at $k = N/2$, in which case $\sum_{j=1}^{N}\theta_{j}  = 0 $.

%\bibliography{cIFMnoisereferences}

%

\end{document}